\documentclass[copyright,creativecommons]{eptcs}

\providecommand{\event}{M. Bujorianu and M. Fisher (Eds.):\\ \hbox{\quad}Workshop on Formal Methods for Aerospace (FMA)}
\providecommand{\volume}{20}
\providecommand{\anno}{2010}
\providecommand{\firstpage}{34}
\providecommand{\eid}{4}

\usepackage{abbrevs-gb}
\usepackage{mycommands}
\usepackage{lustre}
\usepackage{mathpartir}
\usepackage{amsmath}
\usepackage{amsthm}
\usepackage{amssymb}
\usepackage{graphicx}
\usepackage{algorithm}
\usepackage{epsfig}
\usepackage{subfigure}
\usepackage{algpseudocode}
\usepackage{pdfpages}

\newtheorem{definition}{Definition}
\newtheorem{property}{Property}

\title{Implementing Multi-Periodic Critical Systems: from
  Design to Code Generation\thanks{This work was funded by \astrium}}

\author{Julien Forget
  \institute{ONERA\\ Toulouse, France}
  \email{~~~julien.forget@onera.fr~~~}
  \and
  Fr\'ed\'eric Boniol
  \institute{ONERA\\ Toulouse, France}
  \email{frederic.boniol@onera.fr}
  \and
  David Lesens
  \institute{EADS Astrium Space Transportation\\Les Mureaux, France}
  \and
  Claire Pagetti
  \institute{ONERA\\ Toulouse, France}
  \email{claire.pagetti@onera.fr}}

\def\titlerunning{Implementing Multi-Periodic Critical Systems: from
  Design to Code Generation}
\def\authorrunning{J. Forget, F. Boniol, D. Lesens \& C. Pagetti}

\begin{document}

\lstset{basicstyle=\ttfamily, language=Lustre}

\maketitle

\begin{abstract}

  This article presents a complete scheme for the development of
  Critical Embedded Systems with Multiple Real-Time Constraints. The
  system is programmed with a language that extends the synchronous
  approach with high-level real-time primitives. It enables to assemble
  in a modular and hierarchical manner several locally mono-periodic
  synchronous systems into a globally multi-periodic synchronous
  system. It also allows to specify flow latency constraints. A program
  is translated into a set of real-time tasks. The generated code (\C\
  code) can be executed on a simple real-time platform with a
  dynamic-priority scheduler (EDF). The compilation process (each
  algorithm of the process, not the compiler itself) is formally proved
  correct, meaning that the generated code respects the real-time
  semantics of the original program (respect of periods, deadlines,
  release dates and precedences) as well as its functional semantics
  (respect of variable consumption).
\end{abstract}



\section{Introduction}
Embedded systems have successfully been implemented with synchronous
languages in the past. In particular, data-flow synchronous languages
(\lustre / \scade\ \cite{halbwachs91a}, \signal\ \cite{benveniste91})
are well adapted for describing precisely the data flow between the
communicating processes of the system. In \cite{forget08c} we proposed
an extension of synchronous languages to design multi-periodic systems
efficiently, by assembling several synchronous nodes into a
multi-periodic synchronous program. Such an approach allows to describe
the real-time aspects and the functional aspects of a system in the same
framework. The purpose of the paper is to give an overview of the
language capabilities and to describe the compilation chain through the
programming of a case study. The generated code is targeted for
a simple real-time platform with the \textit{earliest-deadline-first}
(EDF) scheduling policy \cite{liu73}. We present the whole
compilation chain but we do not get into the details of the proofs,
which can be found in \cite{forget09bgb}. We focus more particularly on
the generated code, which gives a concrete illustration of the
compilation and summarizes our contribution.

\subsection{Motivation}

The development of an industrial critical system may involve several
teams, which separately define the different functions of the
system. The functions are then assembled by the integrator, who
implements the communications between the functions. Currently, this
integration process lacks a formal language to ease the design process
and to ensure the correctness of the global system.

\begin{figure}[hbt]
  \centering \includegraphics[width=.8\linewidth]{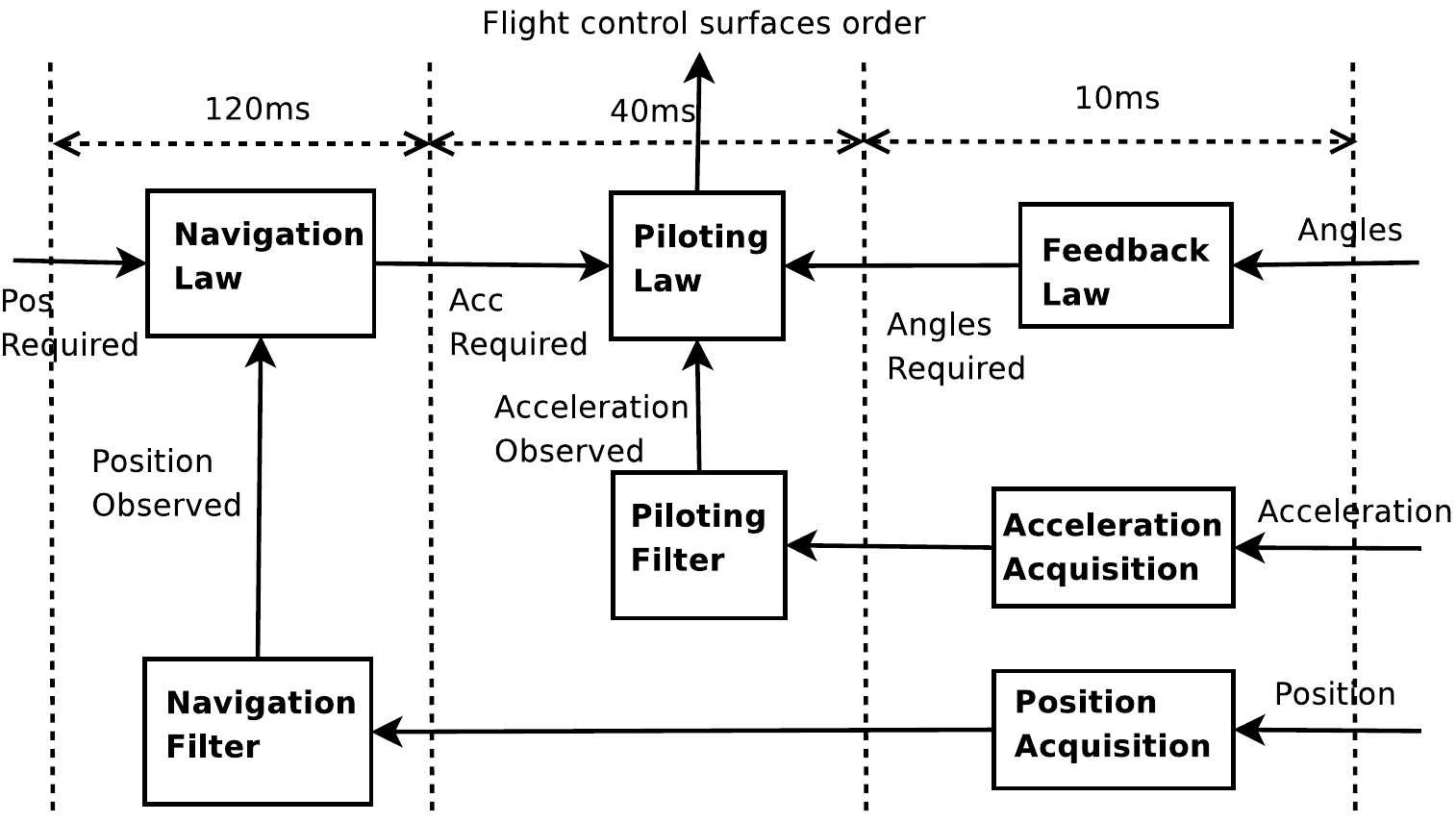}
  \caption{Flight control system}
  \label{fig:CDV}
\end{figure}

We consider the simplified Flight Control System of Fig. \ref{fig:CDV}
as a case study. This system controls the attitude, the trajectory and
the speed of an airplane in auto-pilot mode. It consists of three
communicating sub-systems. Each sub-system consists of several
operations (represented by boxes in the figure) and executes repeatedly
at a periodic rate. The fastest sub-system executes at 10ms, it acquires
the state of the system (angles, position, acceleration) and computes
the feedback law of the system. The intermediate sub-system is the
piloting loop, it executes at 40ms and manages the flight control
surfaces of the airplane. The slowest sub-system is the navigation loop,
it executes at 120ms and determines the acceleration to apply. The
required position of the airplane is acquired at the slow rate.

The three sub-systems are first defined separately by different
teams. The integrator then assembles them in the global system and
specifies how they communicate. The language focuses on this assembly
level.

\subsection{Contribution}
The main novelties of our approach are: first the integrator can develop
the complete system in a unified formal framework (a high-level
formal language) and second the language along with its compiler covers
the development of the system from its design to its implementation,
through automated code generation.

This relies on two different research domains. On the one hand,
scheduling theory focuses mainly on satisfying system real-time
constraints but usually disregards system functional behaviour. This
ensures the correctness of the temporal properties of the system, but
makes it hard to verify the functional correctness of the system. In the
case of multi-periodic systems, this often leads to non-deterministic
process communications. On the other hand, synchronous languages focus
on the correctness of the functional behaviour of the system and ensure
that it is deterministic. However, classic synchronous languages
abstract from real-time (except some recent extensions discussed in
Sect. \ref{sec:related}), which makes them ill-adapted to the
implementation of multi-periodic systems.

Our work combines scheduling theory and synchronous languages to ensure
both the functional and the temporal correctness of multi-periodic
systems. This is particularly suitable for the implementation of
critical systems. The integrator programs the system with the language
introduced in \cite{forget08c}, which extends synchronous languages with
high-level real-time primitives. The compiler then generates the set of
real-time tasks corresponding to this program. Tasks are then translated
into \C\ threads, each one containing the functional code of a task
completed with a deterministic data-exchange protocol that does not
require synchronization primitives (such as semaphores). The threads are
scheduled concurrently with the EDF policy. They can be preempted by the
scheduler during their execution but preemptions do not jeopardize the
functional correctness of the system. The use of an EDF scheduler
departs from the classic compilation scheme of synchronous languages,
which translates a program into a ``single-loop'' sequential code
\cite{halbwachs91b}. The single-loop scheme relies on a static-priority
non-preemptive scheduling policy, which makes it ill-adapted for
implementing multi-periodic processes. A dynamic-priority preemptive
policy like EDF allows to achieve better processor utilization (to
execute more time-consuming processes). The complete compilation process
has been implemented in \ocaml\ and is about 3000 code lines long. It
generates \C\ code with calls to the real-time primitives defined in the
real-time extensions of POSIX \cite{posix13}.

\subsection{Paper Outline}
Sect. \ref{sec:language} gives an overview of the language for
specifying multi-periodic systems and shows how the case study can be
programmed. We then detail the compilation process. The correctness
of the system is first verified by a series of static analyses
(Sect. \ref{sec:static-analysis}). We then translate the program into a
set of real-time tasks (Sect. \ref{sec:task-translation}). The
preservation of the synchronous semantics during inter-task
communications is ensured by a buffering protocol described in
Sect. \ref{sec:communications}. We can then translate the tasks into \C\
code for a simple real-time platform (Sect. \ref{sec:implementation}). A
comparison with related works is given in Sect. \ref{sec:related}.

\section{A Synchronous Real-Time Language}
\label{sec:language}
\subsection{Informal Presentation}
\label{sec:CDV-prog}
We present the language through the programming of the Flight Control
System of Fig. \ref{fig:CDV}. The different operations of the Flight
Control System are first declared as imported nodes, named after the
initials of the operations (for instance, \lstinline!PA!  stands for
``Position Acquisition''):
\begin{lstlisting}[basicstyle=\scriptsize]
imported node PA(i: int) returns (o: int) wcet 1;
imported node AA(i: int) returns (o: int) wcet 1;
...
\end{lstlisting}
The declaration of an imported node specifies the inputs and outputs of
the node with their types and the worst case execution time (wcet) of the
node. For instance, the node \lstinline!PA! has one input \lstinline!i!
of type \lstinline!int!, one output \lstinline!o! of type
\lstinline!int! and its wcet is 1.

For each sub-system, we define an intermediate node that groups the
operations of the sub-system. Node definitions are modular and
hierarchical. There are several ways to decompose the Flight Control
System into nodes and it is also possible to program the whole system as
a single node. However, the different decompositions produce the same
behaviour as intermediate nodes are flattened during the compilation
(see Sect. \ref{sec:extraction}). We choose to group operations by
sub-systems (and by rates) for better readability. In the following, the
suffix \lstinline!_o! stands for ``observed'', \lstinline!_r! for
``required'' and \lstinline!_i! for ``intermediate''. The node for the
acquisition loop is defined as follows:
\begin{lstlisting}[basicstyle=\scriptsize]
node acquisition(angle,pos,acc)returns(pos_i,acc_i,angle_r)
let
  pos_i = PA(pos);
  acc_i = AA(acc);
  angle_r = FL(angle);
tel
\end{lstlisting}
This node has three inputs and three outputs, the types of which are
left unspecified and will be inferred by the type-checker of the
language (see Sect. \ref{sec:static-analysis}). The body of the node (the
\lstinline!let ... tel! block) is a set of equations that define how the
outputs of the node are computed from its inputs. All the variables and
expressions of a program are flows. For example, the constant value $0$
stands for an infinite constant sequence. Nodes are applied point-wisely
to their arguments. So, for instance, at each repetition of node
\lstinline!acquisition!, the output \lstinline!pos_i! is obtained by
applying node \lstinline!PA! to input \lstinline!pos!.

Similarly, we define a node for the piloting loop and for the
navigation loop:
\begin{lstlisting}[basicstyle=\scriptsize]
node piloting (angle_r, acc_i, acc_r) returns (order)
var acc_o;
let
  acc_o = PF(acc_i);
  order = PL (angle_r, acc_o, acc_r);
tel

node navigation (pos_i, pos_r) returns (acc_r)
var pos_o;
let
  pos_o = NF(pos_i);
  acc_r = NL(pos_o,pos_r);
tel
\end{lstlisting}

So far, each node could be defined with the existing \lustre\ language
as each sub-system is mono-periodic. For the main node \lstinline!FCS!
however, we use new primitives to handle \emph{rate transitions} (when
operations of different rates communicate) and to specify the real-time
constraints of the different operations:
\begin{lstlisting}[basicstyle=\scriptsize]
node FCS (pos_r: rate (120, 0); angle, pos, acc) returns (order: due 15)
var acc_i, acc_r, angle_r, pos_i;
let
  acc_r = navigation(pos_i/^12,pos_r);
  order = piloting(angle_r/^4, acc_i/^4, (0 fby acc_r)*^3);
  (pos_i, acc_i, angle_r) = acquisition(angle,pos,acc);
tel
\end{lstlisting}
When a faster node consumes a flow produced by a slower node, we
under-sample the flow using operator $\Dclock$. $e \Dclock k$ only keeps
the first value out of each $k$ successive values of $e$. For instance
flow \lstinline!acc_i! is under-sampled by factor 4 as its consumer
(\lstinline!piloting!) is 4 times slower than its producer
(\lstinline!acquisition!).

For communications from slow to fast operations, we first delay the flow
with operator $\Fbyshort$. The operator $\Fbyshort$ inserts a unitary
delay: expression $cst\Fby e$ produces the value $cst$ at its first
iteration and then the previous values of $e$ (ie $e$ delayed by the
period of $e$). We then over-sample the delayed flow with operator
$\Uclock$. $e\Uclock k$ over-samples $e$ by a factor $k$. Each value of
$e$ is duplicated $k$ times in the result. For instance the flow
\lstinline!acc_r! is delayed and then over-sampled by a factor 3 as its
consumer (\lstinline!piloting!) is 3 times faster than its producer
(\lstinline!navigation!). We use a delay before over-sampling the flow
to avoid reducing the deadline for the production of the
flow. In the case of flow \lstinline!acc_r! for instance, without a
delay the deadline for \lstinline!NL!  would be lower than 40.

For now, we have only described the ratio between the execution rates of
the nodes. The declaration of the node inputs simply specifies that
\lstinline!pos_r! has clock $\isp{120,0}$ (ie that it has period 120 and
phase 0) and all the different rates of the system are deduced from this
information by the clock calculus (see
Sect. \ref{sec:static-analysis}). The declaration of output
\lstinline!order!, imposes a deadline constraint (\lstinline!due 15!),
which requires it to be produced less than 15ms after the beginning of
its period, to respect some external environment constraint. Its period is
left unspecified (its inferred value is 40ms). The behaviour of the new
operators is illustrated in Fig. \ref{fig:ex-synchronous}, in which we
give the value of each expression at each repetition of the system.

\begin{figure}[hbt]
  \centering
{\small
  \begin{tabular}{l|cccccc}
    \hline
    date & 0 & 10 & 20 & 30 & 40 & \ldots \\
    \hline
    \lstinline!angle_r! & $an_0$ & $an_1$ & $an_2$ & $an_3$ & $an_4$ & \ldots \\
    \hline
    \lstinline!angle_r/^4! & $an_0$ & & & & $an_4$ & \ldots\\
    \hline
    \hline
    date & 0 & 40 & 80 & 120 & 160 & \ldots\\
    \hline
    \lstinline!acc_r! & $ac_0$ & & & $ac_1$ & & \ldots\\
    \hline
    \lstinline!0 fby acc_r! & 0 & & & $ac_0$ & & \ldots\\
    \hline
    \lstinline!(0 fby acc_r)*^3! & 0 & 0 & 0 & $ac_0$ & $ac_0$ & \ldots\\
    \hline
  \end{tabular}
}
  \caption{Behaviour of real-time operators}
  \label{fig:ex-synchronous}
\end{figure}


\subsection{Formal Definition: \PCLOCKS}
In the synchronous approach, the computations performed by the system
are split into a succession of \emph{instants}, where each instant
corresponds to one repetition of the system. The synchronous assumption
requires that for each instant, computations end before the end of the
instant. Computations can be activated or deactivated at different
instants using \emph{clocks}. Clocks define the temporal behaviour of
the program on the logical time scale of instants.

To define formally the real-time operators presented in the previous
section, we introduce a new class of clocks called
\emph{\pclocks}. Given a set of \emph{values} $\mathcal{V}$, a
\emph{flow} is a sequence of pairs $(v_i,t_i)_{i\in\mathbb{N}}$ where
$v_i$ is a value in $\mathcal{V}$ and $t_i$ is a tag in $\mathbb{N}$,
such that for all $i$, $t_i<t_{i+1}$. The clock of a flow is its
projection on $\mathbb{N}$. A tag represents an amount of time elapsed
since the beginning of the execution of the program. Each value of a
flow must be computed before its next activation: $v_i$ must be produced
during the time interval $[t_i,t_{i+1}[$. After precedence encoding, the
deadline may actually be less than $t_{i+1}$, this will be detailed in
Sect. \ref{sec:deadline-calculus}.

\begin{definition} (\emph{\Pclock}).  A clock
  $h=(t_i)_{i\in\mathbb{N}}$, $t_i\in\mathbb{N}$, is \emph{strictly
    periodic} if and only if: $\exists n\in\mathbb{N}^{*},\;\forall
  i\in\mathbb{N},\; t_{i+1}-t_i=n$.

  $n$ is the \emph{period} of $h$,
  denoted $\Period{h}$ and $t_0$ is the \emph{phase} of $h$, denoted
  $\Phaseof{h}$.
\end{definition}
\begin{definition}
  The term $\isp{n,p}\in\mathbb{N}^{*}\times\mathbb{Q}^+$ denotes the
  \pclock\ $\alpha$ such that $\Period{\alpha} = n$ and $\Phaseof{\alpha} = \Period{\alpha}*p$
\end{definition}

A \pclock\ defines the real-time rate of a flow and is uniquely
characterized by its phase and by its period. \Pclocks\ relate logical
time (instants) to real-time. Locally, each flow has its own notion of
instant (it must end before its next activation), and globally we can
compare flows that do not share the same notion of instants by relating
instants to real-time. We introduce \transps\ to formalize such rate
transitions:


\begin{definition}
  \label{def:transfs}
  Let $\alpha$ be a \pclock, operations $\Each,\Times$ and $\Phase$ are \transps,
  that produce new \pclocks\ satisfying the following properties:
  \begin{itemize}
  \item $\Period{\alpha \Each k} = k*\Period{\alpha},\;\Phaseof{\alpha
      \Each k} = \Phaseof{\alpha},k\in\mathbb{N}^{*}$
  \item $\Period{\alpha \Times k} = \Period{\alpha}/k,\;\Phaseof{\alpha
      \Times k} = \Phaseof{\alpha}, k\in\mathbb{N}^{*}$
  \item $\Period{\alpha \Phase q} = \Period{\alpha},\;\Phaseof{\alpha \Phase
      q} = \Phaseof{\alpha}+q*\Period{\alpha}, q\in \mathbb{Q}$
  \end{itemize}
\end{definition}
Rate transition operators apply \transps\ to flows. If $e$
has clock $\alpha$, then $e\Dclock k$ has clock $\alpha\Each k$,
$e\Uclock k$ has clock $\alpha\Times k$ and $e\Phclock q$ has clock
$\alpha\Phase q$.



\subsection{Syntax}
\label{sec:syntax}
The syntax of the language is close to \lustre. It is extended with
real-time primitives based on \pclocks. The grammar of the language is
given in Fig. \ref{fig:grammar}. A program consists of a list of node
declarations ($nd$). Nodes can either be defined in the program
($\Nodeshort$) or implemented outside ($\ImportedNodeshort$), for
instance by a \C\ function. Node durations must be provided for each
imported node, more precisely an upper bound on worst case execution
times ($wcet$). The external code provided for imported nodes can itself
be generated by a standard synchronous language compiler (like \lustre),
in case developers want to program the whole system using synchronous
languages. The clock of input/output parameters ($in/out$) of a node can
be declared strictly periodic ($x:\Tperiodic(n,p)$, $x$ then has clock
$\isp{n,p}$) or unspecified ($x$). A deadline constraint can be imposed
on outputs ($x: \Tperiodic(n,p)\due n'$, the deadline is $n'$). The body
of a node consists of an optional list of local variables ($var$) and a
list of equations ($eq$). Each equation defines the value of one or
several variables using an expression on flows ($var=e$). Expressions
may be immediate constants ($cst$), variables ($x$), pairs ($(e,e)$),
initialised delays ($cst\Fby e$), applications ($N(e)$) or expressions
using \pclocks\ ($epck$). Values $k$, $n$, $n'$, $q$ must be statically
evaluable. Value $q$ must be an element of $\mathbb{Q}^+$.

\begin{figure}[htb]
  \centering
  \[
\begin{array}{lcl}
  cst & ::= & \True \Alt \False \Alt 0 \Alt ...\\
  var & ::= & x\Alt var,var\\
  e & ::= & cst\Alt x \Alt (e,e) \Alt cst\Fby e\Alt N(e) \Alt epck\\
  epck & ::= & e\Dclock k \Alt e\Uclock k\Alt e\Phclock q\\
  eq & ::= & var=e \Alt eq;eq \\
  in & ::= & x:\Tperiodic(n,p) \Alt x \Alt in;in\\
  out & ::= & x:\Tperiodic(n,p) \Alt x: \Tperiodic(n,p)\due n'\Alt x\Alt out;out\\
  nd & ::= & \Node{N}{in}{out}[\Var var;] {\Let{eq}}\\
  & & \Alt \kword{imported}\kword{node} N(in)\kword{returns}(out)\wcet{n};
\end{array}
\]
  \caption{Language grammar}
  \label{fig:grammar}
\end{figure}


\section{Static Analyses}
\label{sec:static-analysis}
Synchronous languages are targeted for critical systems. Therefore, the
compilation process puts strong emphasis on the verification of the
correctness of the programs to compile. This consists of a series of
static analyses of the program, which are performed before code
generation.

The first analysis performed by the compiler is the type-checking. The
language is a strongly typed language, in the sense that the execution
of a program cannot produce a run-time type error. Each flow has a
single, well-defined type and only flows of the same type can be
combined. The type-checking of the language is fairly standard
\cite{pierce02}. For the example of the Flight Control System, the
type-checker produces the flowing type for node \lstinline!FCS!:
\lstinline!(int*int*int*int)->int!. This means that the node takes four
integer inputs and produces one integer output. This type is inferred
from the types of the imported nodes.

The causality check verifies that the program does not contain cyclic
definitions: a variable cannot instantaneously depend on itself (i.e.\
not without a \lstinline!fby! in the dependencies). For instance, the
equation \lstinline!x=x+1;! is incorrect, it is similar to a deadlock
since we need to evaluate \lstinline!x+1! to evaluate \lstinline!x! and
we need to evaluate \lstinline!x! to evaluate \lstinline!x+1!.

The clock calculus (defined in \cite{forget08c}) verifies that a program
only combines flows that have the same clock. When two flows have the
same clock, they are \textit{synchronous} as they are always present at
the same instants. Combining non-synchronous flows leads to
non-deterministic programs as we access to undefined values. For
instance we can only compute the sum of two synchronous flow, because
the value of the sum is ill-defined when only one of the two flows is
absent (when the two flows are absent the sum flow is simply absent,
which is well-defined). The clock calculus ensures that a synchronous
program will never access to undefined values. For the example of the
Flight Control System, the clock-calculus produces the following clock
for node \lstinline!FCS!:
\lstinline!((120,0)*(10,0)*(10,0)*(10,0))->(40,0)!. This means that
inputs \lstinline!angle!, \lstinline!acc!, \lstinline!position! have
period 10, while input \lstinline!position_r! has period 120. The output
\lstinline!order! has period 40 (though its deadline is 10). These
static analyses ensure that a program accepted by the compiler has a
deterministic behaviour.


\section{Translation into Real-Time Tasks}
\label{sec:task-translation}

This section details how the compilation process translates the input
program into a set of real-time tasks. We first extract a task graph
from the program, where tasks are related by precedence constraints. We
then encode task precedences in task real-time attributes to obtain a
set of independent tasks.

\subsection{Task Graph Extraction}
\label{sec:extraction}
\subsubsection{Tasks}

A synchronous program consists of a hierarchy of nodes, the leaves of
which are predefined or imported nodes. The task graph generation
process first inlines intermediate nodes appearing in the main node
recursively, replacing each intermediate node call by its set of
equations. For instance, the program of the Flight Control System of
Sect. \ref{sec:CDV-prog} is translated into a single node (the main node
\lstinline!FCS!) containing one node call to each imported node,
\lstinline!PA!, \lstinline!AA!, \lstinline!FL!,
\lstinline!PF!, \lstinline!PL!, \lstinline!NF!, \lstinline!NL!, one node
call to operator $\Uclock$, one node call to operator \lstinline!fby!
and three node calls to operator $\Dclock$.

This ``flattened'' main node is then translated into a task graph. Each
imported node call is translated into a task. Each variable of the node
and predefined operator call is also translated into a vertex but will
later be reduced to simplify the graph (see
Sect. \ref{sec:reduced-graph}). The clustering of several nodes into the
same task to reduce the number of generated tasks is
beyond the scope of this paper. We could probably reuse existing
strategies, for instance those suggested in \cite{curic05}.

\subsubsection{Task Precedences}
In order to respect the synchronous semantics, for each data-dependency there
must be a precedence from the task producing the data to the task
consuming it. Task precedences are deduced from data dependencies
between expressions of the program. Similarly to \cite{halbwachs91b}, we
say that an expression $e'$ precedes an expression $e$ when $e$
syntactically depends on $e'$. This occurs either when $e'$ appears in
$e$ or when $x$ appears in $e$ and we have an equation $x=e$. Let
$g=(V,E)$ denote a task graph, where $V$ is the set of tasks of the
graph and $E$ is the set of task precedences of the graph (a subset of
$V\times V$). For instance, the flattened Flight Control program
contains the two equations \lstinline!pos_o=NF(pos_i/^12);!
\lstinline!pos_i=PA(pos);!. These equations produce the following graph:
$(\{$\lstinline!NF,PA,/^12,pos,pos_i,pos_o!$\}$,
$\{$\lstinline!pos!$\precrel$ \lstinline!PA!, \lstinline!PA!$\precrel$
\lstinline!pos_i!, \lstinline!pos_i!$\precrel$ \lstinline!/^12!,
\lstinline!/^12!$\precrel$ \lstinline!NF!, \lstinline!NF!  $\precrel$
\lstinline!pos_o!$\})$.

\subsubsection{Task Graph Reduction}
\label{sec:reduced-graph}
We then simplify the intermediate graph structure. First, each input of
the main node is translated into a (sensor) task and each output of the
main node is translated into a (actuator) task. Second, we remove
variables from the graph, replacing recursively each pair of precedence
$N\precrel x\precrel M$, where $x$ is a local variable, by a single
precedence $N\precrel M$.

\begin{figure}[hbt]
  \centering
  \centering \includegraphics[width=.5\linewidth]{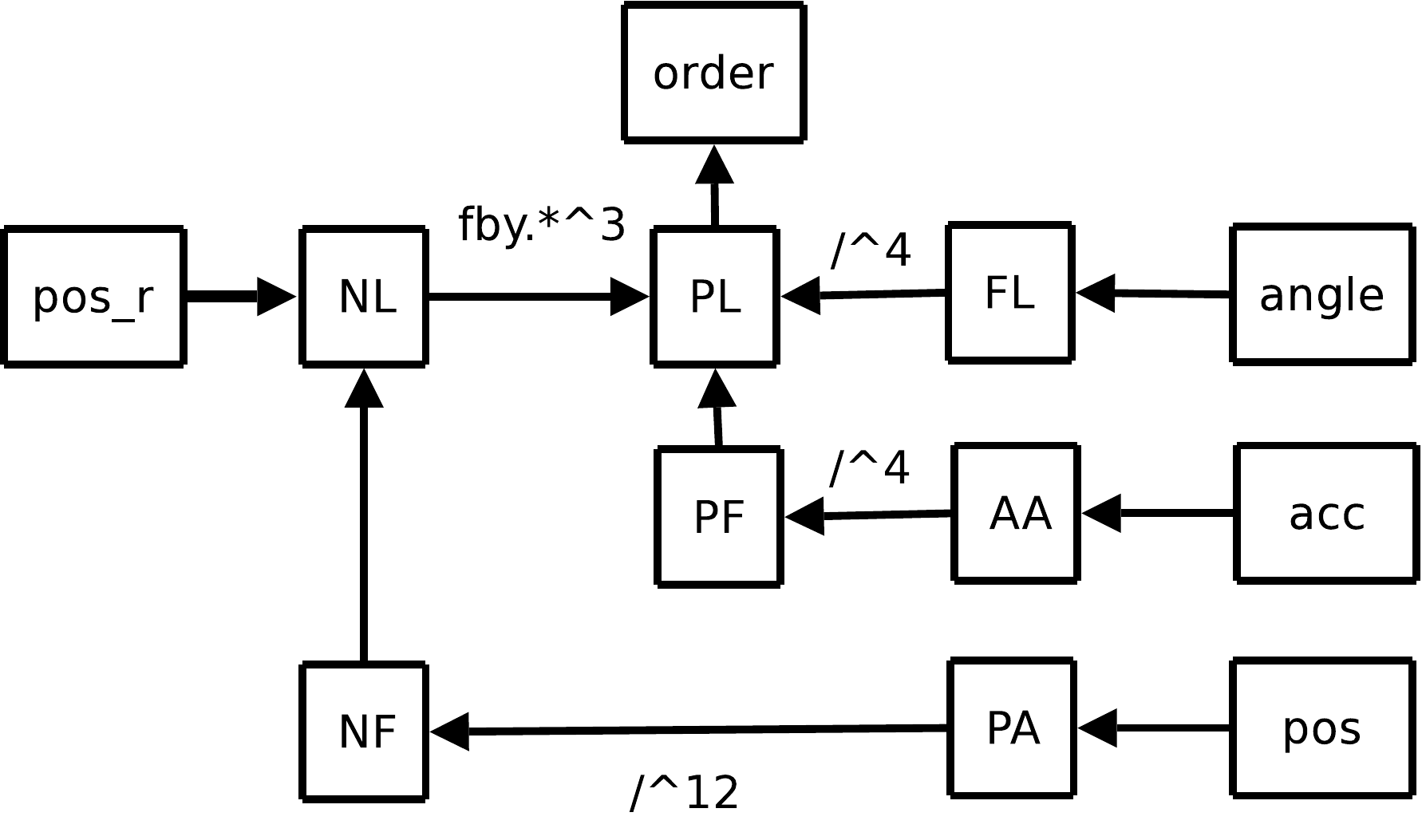}
  \caption{Reduced task graph for the Flight Control System program}
  \label{fig:prec-graph}
\end{figure}

We finally translate predefined nodes into \textit{precedence
  annotations}. A precedence $\tau_i\precany\tau_j$ represents an
\textit{extended precedence}, where $ops$ is a list of precedence
annotations $op$, with $op\in\{\Uclock k,\Dclock k,\Phclock
q,\Fbyshort\}$. $\List{op}{ops}$ denotes the list whose head is $op$ and
whose tail is $ops$ and $\epsilon$ denotes the empty list. For instance,
the precedences \lstinline!PA!$\precrel$ \lstinline!pos_i!,
\lstinline!pos_i!$\precrel$ \lstinline!/^12!, \lstinline!/^12!$\precrel$
\lstinline!NF! are simplified into a single extended precedence
\lstinline!PA!$\precannot{\Dclock 12}$\lstinline!NF!. When the rewriting terminates,
every task of the graph corresponds to either an imported node or a
sensor or an actuator. The reduced task graph of the Flight Control
System is given in Fig. \ref{fig:prec-graph}.

\subsection{Real-Time Characteristics Extraction}
Each task $\tau_i$ of the graph is characterized by its
real-time attributes $(T_i,C_i,r_i,d_i)$. $\tau_i$ is instantiated
periodically with period $T_i$. It cannot start its execution before all
its predecessors, defined by the precedence constraints, complete their
execution. $C_i$ is the (worst case) execution time of the task. $r_i$
is the release date of the first instance of the task. The subsequent
release dates are $r_i+T_i$, $r_i+2T_i$, etc. $d_i$ is the relative
deadline of the task. The absolute deadline $D_i[j]$ of the instance $j$
of a task $\tau_i$ is the release date $R_i[j]$ of this instance plus
the relative deadline of the task: $D_i[j]=R_i[j]+d_i$.  Task real-time
characteristics are extracted as follows:
\begin{itemize}
\item \emph{Periods}: The period of a task is obtained from its clock $ck_i$. We have $T_i=\Period{ck_i}$.
\item \emph{Deadlines}: By default, the deadline of a task is its period ($d_i=T_i$). Deadline
constraints can also be specified on the production of a node output
(\lstinline!o: due n!).
\item \emph{Release Dates}: The initial release date of a task is the phase of its clock:
$r_i=\Phaseof{ck_i}$.
\item \emph{Execution Times}: The execution time $C_i$ of a task is
  directly specified by the \lstinline!wcet! of the imported node
  declaration. For simplification, we consider that the run-time
  overhead due to task preemptions is negligible.
\end{itemize}



\subsection{Precedence Encoding}
\label{sec:prec-encoding}
\subsubsection{Simple Precedences Encoding}
\cite{chetto90} showed that a set of dependent tasks (task related by
precedence constraints) can be reduced to a set of independent ones
(without precedences) obtaining an equivalent problem under the EDF
policy, by adjusting task release dates and deadlines such that
precedences are encoded in the adjusted real-time characteristics. The
adjusted absolute deadline $D^*_i$ of a task is:
\[D^*_i=\min_{\tau_j\in succ(\tau_i)}(D_i,min(D^*_j-C_j))\]
If we want to perform a schedulability test, the adjusted release date of a task
is: 
\[R^*_i=\max_{\tau_j\in pred(\tau_i)}(R_i,max(R^*_j+C_j))\]
 If we
only want to schedule the program correctly, the adjusted release date
of a task $R^{*'}_i$ is:
\[R^{*'}_i=\max_{\tau_j\in pred(\tau_i)}(R_i,max(R^{*'}_j))\]
For simplification, we only
consider the second encoding in the following.

\subsubsection{Extended Precedences Encoding}
\begin{figure}[hbt]
  \centering
  \subfigure[$A\precdiv{2}B$]
  {\label{fig:fast-to-slow-prec}
    \includegraphics[width=.2\linewidth]{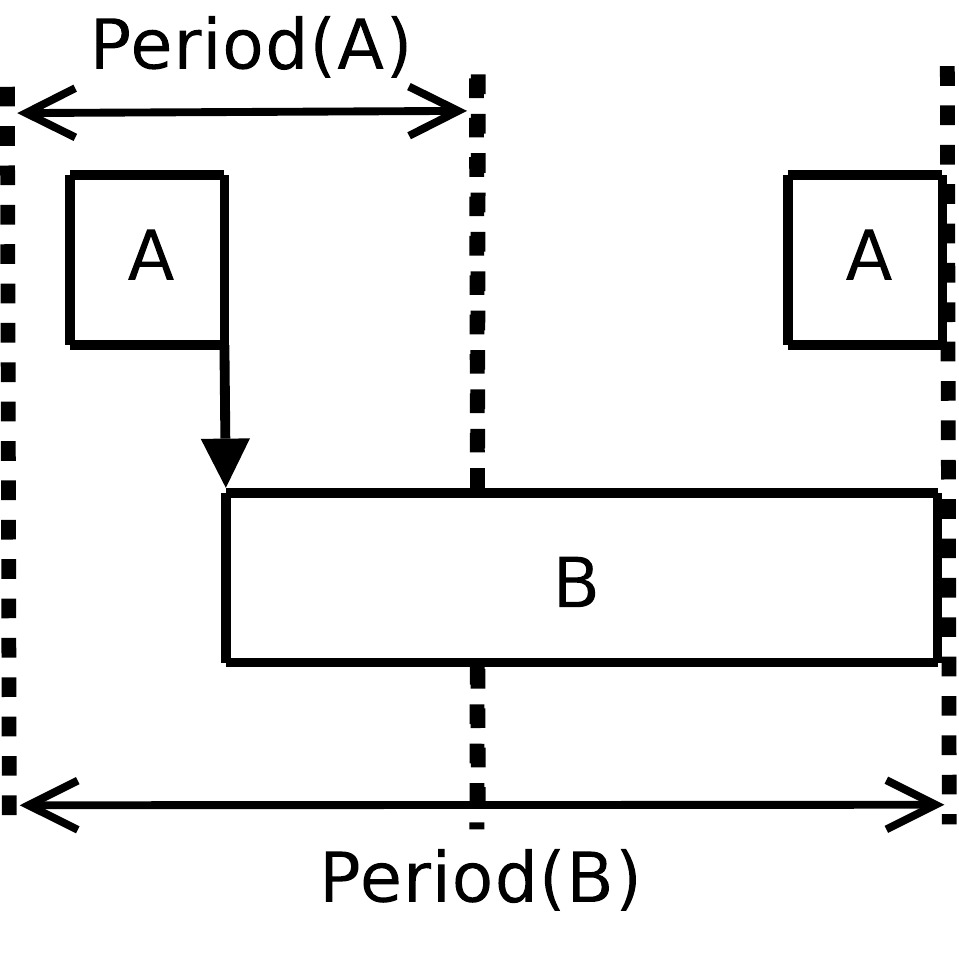}}
  \hfill
  \subfigure[$A\precmult{2}B$]
  {\label{fig:slow-to-fast-prec}
    \includegraphics[width=.2\linewidth]{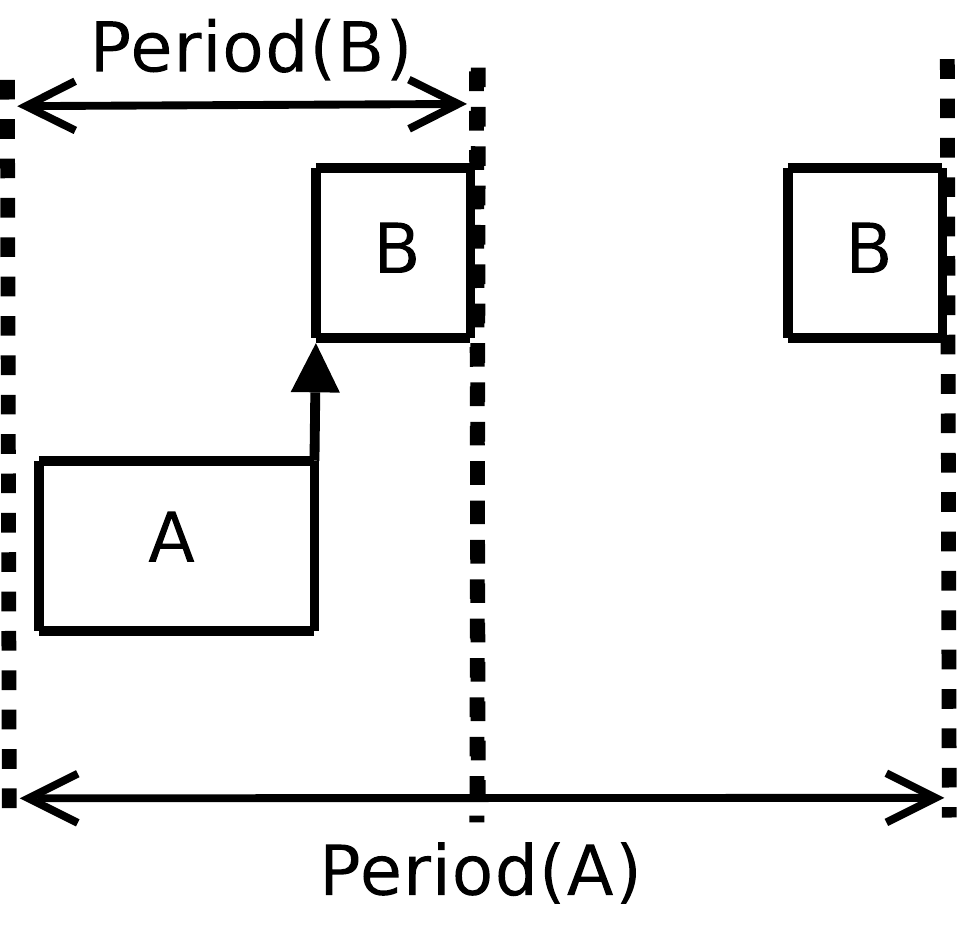}}
  \hfill
  \subfigure[$A\precdelay B$]
  {\label{fig:delayed-prec}
    \includegraphics[width=.2\linewidth]{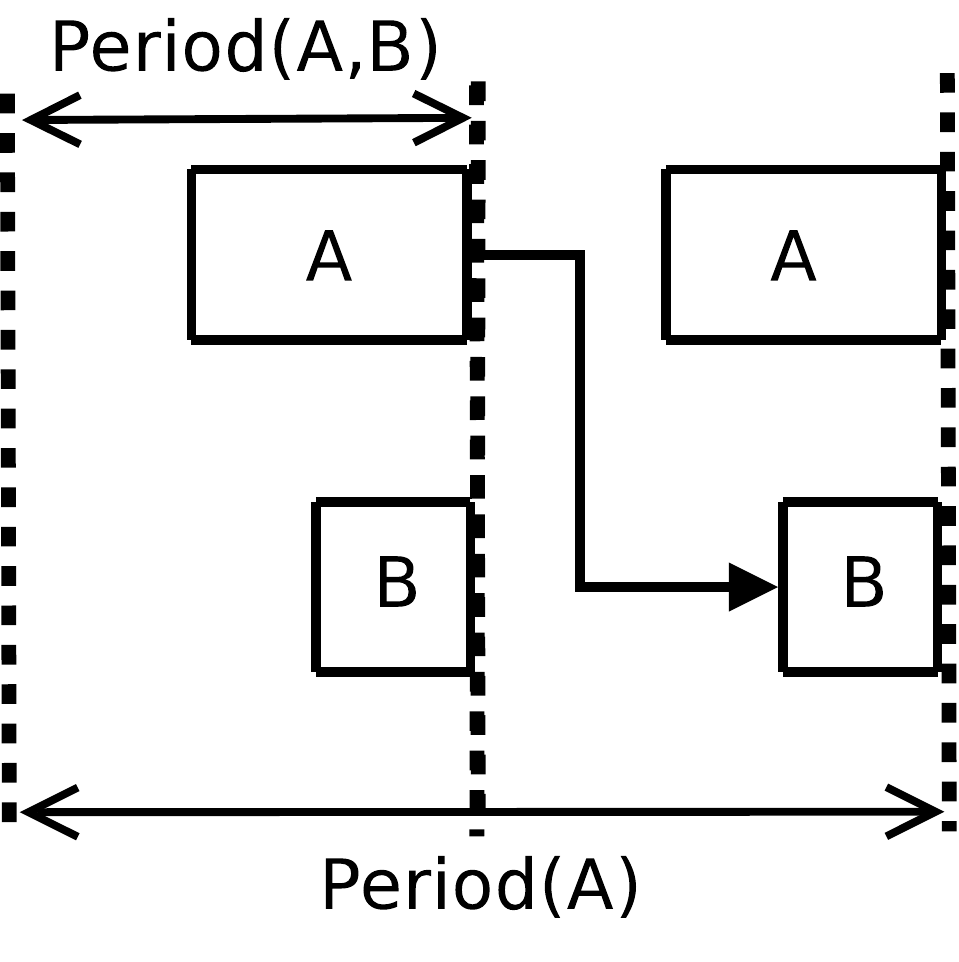}}
  \hfill
  \subfigure[$A\precoffset{1/2}B$]
  {\label{fig:offset-prec}
    \includegraphics[width=.2\linewidth]{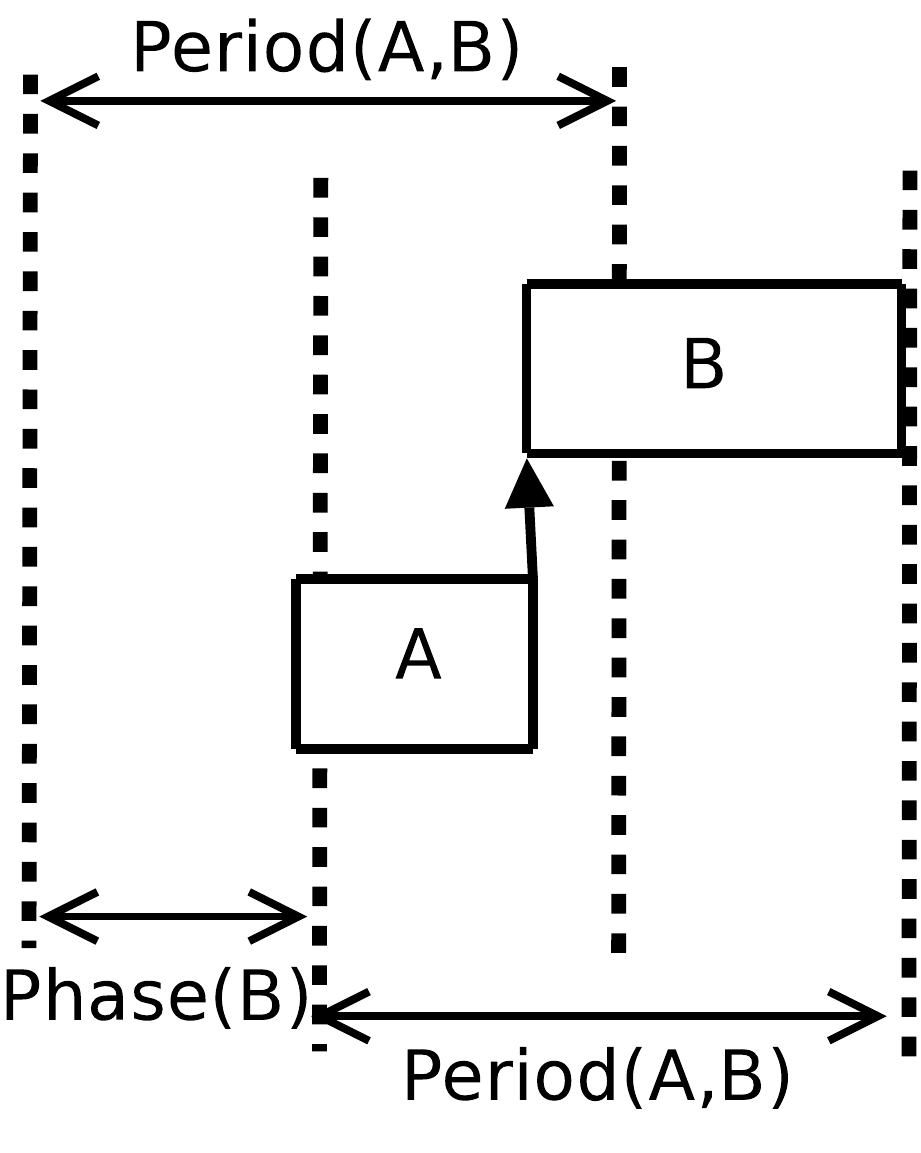}}
  \caption{Encoding extended precedences}
  \label{fig:multi-prec}
\end{figure}

Fig. \ref{fig:multi-prec}, shows that we can unfold extended precedences
between tasks into simple precedences between task instances. The
precedence encoding technique can then be applied to
the ``unfolded'' graph.

More formally, let $\tau_i[n]\precrel\tau_j[n']$ denote a precedence
from task instance $\tau_i[n]$ to task instance $\tau_j[n']$. From the
semantics of predefined operators, we have
$\tau_i\precany\tau_j\Rightarrow \forall n,\;\tau_i[n]\precrel
\tau_j[g_{ops}(n)]$, with $g_{ops}$ defined as follows:
\begin{align*}
  &g_{\List{\Uclock k}{ops}}(n) = g_{ops}(kn) & g_{\List{\Dclock
      k}{ops}}(n) = g_{ops}(\lceil n/k\rceil) \\
  &g_{\List{\Phclock q}{ops}}(n)= g_{ops}(n) & g_{\List{\tiny\Fbyshort}{ops}}(n) = g_{ops}(n+1)\\
  &g_{\Emptylist}(n)= n
\end{align*}
The precedence relation is an over-approximation of the data-dependency
relation. Indeed, there is a data dependency between $\tau_i[n]$ and
$\tau_j[n']$, meaning that $\tau_j[n']$ consumes the data produced by
$\tau_i[n]$, if and only if $\tau_i\precany\tau_j\wedge
n'=g_{ops}(n)\wedge g_{ops}(n)\neq g_{ops}(n+1)$.

We can then adapt the encoding to our context. For each precedence
$\tau_i\precany\tau_j$, we must adjust the release dates and deadlines
of each instance $\tau_i[n]$ such that $R^*_i[n]\leq R^*_j[g_{ops}(n)]$
and $D^*_i[n]\leq D^*_j[g_{ops}(n)]-C_j$. Concerning release dates, we
can easily prove that thanks to the synchronous semantics we already
have $R_i[n]\leq R_j[g_{ops}(n)]$, so release dates do not need to be
adjusted. Concerning deadlines, we need to transpose the formulae to
relative deadlines to fit our task model. From the definition of
relative deadlines: $D^*_i[n]\leq D^*_j[g_{ops}(n)]-C_j\Leftrightarrow
d_i[n]\leq d_j[g_{ops}(n)]+r_j+g_{ops}(n)T_j-r_i-nT_i-C_j$.

\subsubsection{Deadline Calculus}
\label{sec:deadline-calculus}
In practice we do not need to unfold extended precedences to perform
their encoding. Instead, we represent the sequence of deadlines of the
instances of a task as a finite repetitive pattern called
\textit{\dword}. A \emph{unitary deadline} specifies the relative
deadline for the computation of a single instance of a task. It is
simply an integer value $d$. A \emph{\dword} defines the sequence of
unitary deadlines for each instance of a task. The set of \dwords\ is
defined by the following grammar: $w ::=\drepeat{u}\;\;u ::= d \Alt
d.u$. Term $\drepeat{u}$ denotes the infinite repetition of word $u$. In the
following, $w[n]$ denotes the $n^{th}$ unitary deadline of \dword\ $w$
($n\in\mathbb{N}$).

Let $w_i$ denote the \dword\ of task $\tau_i$. A precedence
$\tau_i\precany\tau_j$ is encoded by a constraint relating $w_i$ to
$w_j$ of the form:
\[w_i\leq\wgops_{ops}(w_j)+\diffops_{ops}(T_i,T_j)-C_j+r_j-r_i\] where, for
all $n$, $\wgops_{ops}(w_j)[n]=w_j[g_{ops}(n)]$ and
$\diffops_{ops}(T_i,T_j)[n]=g_{ops}(n)T_j-nT_i$.

Let
$\cstr_{ops}(\tau_j)=\wgops_{ops}(w_j)+\diffops_{ops}(T_i,T_j)-C_j+r_j-r_i$.
\begin{property}
  $\diffops_{ops}(T_i,T_j)$ is periodic and can be represented as a
  \dword. The set of \dwords\ is closed under operation $\wgops_{ops}$
  and under \dwords\ addition. As a consequence, $\cstr_{ops}(\tau_j)$
  is a \dword.
\end{property}
\begin{proof}
  By induction.
\end{proof}

\begin{property}
  The \dwords\ of a task graph $g=(V,E)$ can be computed with complexity
  $\mathcal{O}(|V|+|E|*|w_{max}|)$, where $w_{max}$ denotes the \dword\
  which has the longest size in the task graph.
\end{property}
\begin{proof}
  A reduced task graph is a DAG (when we do not consider delayed
  precedences), so we can compute the \dwords\ of the graph by
  performing a topological sort working backwards
  (starting from the end of the graph). As the complexity of a
  topological sort is $\mathcal{O}(|V|+|E|)$, the complexity of the
  algorithm is $\mathcal{O}(|V|+|E|*|w_{max}|)$ where $w_{max}$ denotes
  the longest \dword\ in the task graph.
\end{proof}

For instance, for the Flight Control System program, we take $C_{PA}=1$,
$C_{AA}=1$, $C_{FL}=3$, $C_{PF}=4$, $C_{PL}=6$, $C_{NL}=20$, $C_{NF}=5$.
To simplify, we take null durations for sensors and actuators. The
result of the deadline calculus is:
$w_{PA}=\drepeat{10}$, $w_{AA}=\drepeat{5.10.10.10}$,
$w_{FL}=\drepeat{9.10.10.10}$, $w_{PF}=\drepeat{9}$,
$w_{PL}=\drepeat{15}$, $w_{NL}=\drepeat{120}$,
$w_{NF}=\drepeat{100}$. The \dwords\ of tasks \lstinline!AA! and
\lstinline!PA! state that, each first repetition out of four successive
repetitions the two tasks have a shorter deadline as \lstinline!PF! and
\lstinline!PL! execute. Notice that if we set deadline $5$ for all the
instances of \lstinline!AA! (instead of a \dword), this example is not
schedulable.


\section{Communication Protocol}
\label{sec:communications}
As task precedences are encoded in task deadlines, inter-task
communications do note require synchronization primitives (like
semaphores for instance). Indeed, as long as tasks respect their
deadlines, data is produced before being consumed, so tasks simply read from
and write to some communication buffers allocated in a global shared
memory when they execute. However, to respect the synchronous semantics,
the input of a task must not change during its execution. Therefore, we
propose a communication scheme, which ensures that the input of a task
remains available until its deadline.

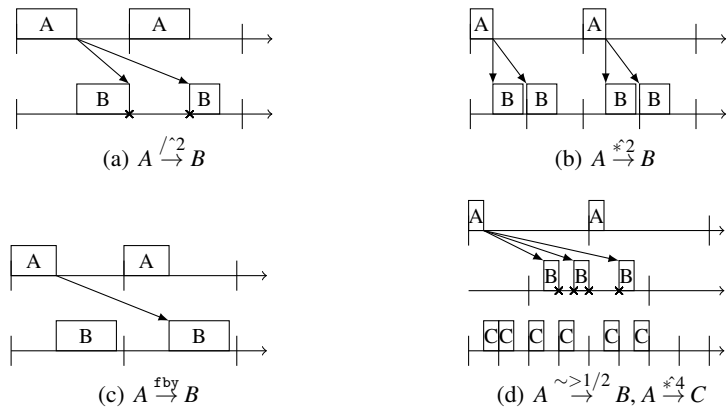
\begin{figure}[hbt]
  \centering
  \subfigure[$A\precdiv{2}B$]
  {\label{fig:fast-to-slow-com}
    \begin{tikzpicture}
      \drawClock{3.4}{0}{1.5}{-0.2}{0.2}{0}{};
      \draw[font=\scriptsize] (0,0.4) rectangle (0.8,0) node[pos=0.5]{A};
      \draw[font=\scriptsize] (1.5,0.4) rectangle (2.3,0) node[pos=0.5]{A};
      \drawClock{3.4}{-1}{3}{-1.2}{-0.8}{0}{};
      \draw[font=\scriptsize] (0.8,-0.6) rectangle (1.5,-1) node[pos=0.5]{B};
      \draw plot[mark=x] (1.5,-1);
      \draw plot[mark=x] (2.3,-1);
      \draw[font=\scriptsize] (2.3,-0.6) rectangle (2.7,-1) node[pos=0.5]{B};
      \draw[-latex] (0.8,0) -- (1.5, -0.6);
      \draw[-latex] (0.8,0) -- (2.3, -0.6);
    \end{tikzpicture}
  }\hspace{2cm}
  \subfigure[$A\precmult{2}B$]
  {\label{fig:slow-to-fast-com}
    \begin{tikzpicture}
      \drawClock{3.4}{0}{1.5}{-0.2}{0.2}{0}{};
      \draw[font=\scriptsize] (0,0.4) rectangle (0.3,0) node[pos=0.5]{A};
      \draw[font=\scriptsize] (1.5,0.4) rectangle (1.8,0) node[pos=0.5]{A};
      \drawClock{3.4}{-1}{0.75}{-1.2}{-0.8}{0}{};
      \draw[font=\scriptsize] (0.3,-0.6) rectangle (0.7,-1) node[pos=0.5]{B};
      \draw[-latex] (0.3,0) -- (0.3, -0.6);
      \draw[font=\scriptsize] (0.75,-0.6) rectangle (1.15,-1) node[pos=0.5]{B};
      \draw[-latex] (0.3,0) -- (0.75, -0.6);
      \draw[font=\scriptsize] (1.8,-0.6) rectangle (2.2,-1) node[pos=0.5]{B};
      \draw[-latex] (1.8,0) -- (1.8, -0.6);
      \draw[font=\scriptsize] (2.25,-0.6) rectangle (2.65,-1) node[pos=0.5]{B};
      \draw[-latex] (1.8,0) -- (2.25, -0.6);
    \end{tikzpicture}
  }\\
  \subfigure[$A\precdelay B$]
  {\label{fig:delayed-com}
    \begin{tikzpicture}
      \drawClock{3.4}{0}{1.5}{-0.2}{0.2}{0}{};
      \draw[font=\scriptsize] (0,0.4) rectangle (0.6,0) node[pos=0.5]{A};
      \draw[font=\scriptsize] (1.5,0.4) rectangle (2.1,0) node[pos=0.5]{A};
      \drawClock{3.4}{-1}{1.5}{-1.2}{-0.8}{0}{};
      \draw[font=\scriptsize] (0.6,-0.6) rectangle (1.4,-1) node[pos=0.5]{B};
      \draw[font=\scriptsize] (2.1,-0.6) rectangle (2.9,-1) node[pos=0.5]{B};
      \draw[-latex] (0.6,0) -- (2.1, -0.6);
    \end{tikzpicture}
  }\hspace{2cm}
  \subfigure[$A\precoffset{1/2}B$, $A\precmult{4}C$]
  {\label{fig:offset-com}
    \begin{tikzpicture}
      \drawClock{3.4}{0}{1.6}{-0.2}{0.2}{0}{};
      \draw[font=\scriptsize] (0,0.4) rectangle (0.2,0) node[pos=0.5]{A};
      \draw[font=\scriptsize] (1.6,0.4) rectangle (1.8,0) node[pos=0.5]{A};
      \drawClock{3.4}{-0.8}{2.4}{-1.}{-0.6}{0.8}{};
      \draw[font=\scriptsize] (1.0,-0.4) rectangle (1.2,-0.8) node[pos=0.5]{B};
      \draw[font=\scriptsize] (1.4,-0.4) rectangle (1.6,-0.8) node[pos=0.5]{B};
      \draw[font=\scriptsize] (2.0,-0.4) rectangle (2.2,-0.8) node[pos=0.5]{B};
      \draw plot[mark=x] (1.2,-0.8);
      \draw plot[mark=x] (1.4,-0.8);
      \draw plot[mark=x] (1.6,-0.8);
      \draw plot[mark=x] (2.0,-0.8);
      \draw[-latex] (0.2,0) -- (1.0,-0.4);
      \draw[-latex] (0.2,0) -- (1.4,-0.4);
      \draw[-latex] (0.2,0) -- (2.0,-0.4);
      \drawClock{3.4}{-1.6}{0.4}{-1.8}{-1.4}{0}{};
      \draw[font=\scriptsize] (0.2,-1.2) rectangle (0.4,-1.6) node[pos=0.5]{C};
      \draw[font=\scriptsize] (0.4,-1.2) rectangle (0.6,-1.6) node[pos=0.5]{C};
      \draw[font=\scriptsize] (0.8,-1.2) rectangle (1.0,-1.6) node[pos=0.5]{C};
      \draw[font=\scriptsize] (1.2,-1.2) rectangle (1.4,-1.6) node[pos=0.5]{C};
      \draw[font=\scriptsize] (1.8,-1.2) rectangle (2.0,-1.6) node[pos=0.5]{C};
      \draw[font=\scriptsize] (2.2,-1.2) rectangle (2.4,-1.6) node[pos=0.5]{C};
    \end{tikzpicture}
  }
  \caption{Communications for extended precedences}
  \label{fig:com-examples}
\end{figure}

For a precedence $\tau_i\precany\tau_j$, data produced by $\tau_i[n]$
may be consumed by $\tau_j[g_{ops}(n)]$ after $\tau_i[n+1]$ has
started. This is illustrated in Fig. \ref{fig:com-examples}, which shows
the schedule of two tasks related by extended precedences. Vertical
lines on the time axis represent task periods and marks represent task
preemptions. An arrow from $A$ at date $t$ to $B$ at date $t'$ means
that task $B$ may read at time $t'$ from the value produced by $A$ at
time $t$. In Fig. \ref{fig:fast-to-slow-com}, \ref{fig:delayed-com} and
\ref{fig:offset-com}, when $A[1]$ executes, it must not overwrite the
data produced by $A[0]$ because it is consumed by $B[0]$ and $B[0]$ is
not complete yet. Therefore, we need a buffer to keep the value of
$A[n]$ after $A[n+1]$ has started. In Fig. \ref{fig:slow-to-fast-com},
the same data is consumed several times but $A[1]$ can freely overwrite
the data produced by $A[0]$, so no specific communication scheme is
required.

The communication protocol allocates a buffer for each extended
precedence of the graph. The producer of the data writes to the buffer
only if $g_{ops}(n)\neq g_{ops}(n+1)$ (see the definition of
data-dependencies in Sect.~\ref{sec:prec-encoding}). The consumer simply
reads from this buffer each time it executes. We allocate a
double-buffer when the precedence contains a $\Fbyshort$ or a
$\Phclock$, to keep the previous and the current value of the data. This
communication scheme is illustrated in more details in
Sect. \ref{sec:task-codegen}. It is of course not optimal because in
many cases when we consider a set of precedences from a single task
$\tau_i$ to several tasks $\tau_j$ we can use the same communication
buffer for some of the tasks $\tau_j$. Optimization will be treated in
future work, we could for instance adapt the communication scheme
proposed by \cite{tripakis05} to our language.


\section{Code Generation}
\label{sec:implementation}
The compiler generates \C\ code with calls to the real-time primitives
defined in the real-time extensions of the POSIX standard (POSIX.13
\cite{posix13}). The code generation can easily be
adapted to any real-time operating system that provides dynamic priority
scheduling. Each task is translated into a thread and the threads are
executed concurrently by an EDF scheduler modified to handle \dwords.

\subsection{Task Code Generation}
\label{sec:task-codegen}
The generated code consists of a single \C\ file. The file starts with
the declaration of one global variable for each communication
buffer. For instance, for the communication from \lstinline!PF! to
\lstinline!PL! (\lstinline!acc_o! before graph expansion), we have the
declaration: \lstinline!int PF_o_PL_i1! (named after the output of the
producer and the input of the consumer). For the communication from
\lstinline!PL! to \lstinline!FL!, we have the declaration:
\lstinline!int PL_o_FL_i2[2]!, as there is a delay between the two
tasks.

The file then contains a function for each task of the task set. The
function mainly consists of an infinite loop, that wraps the function of
the corresponding imported node with the code of the communication
protocol. One step of the loop corresponds to the execution of one
instance of the task. Once buffer updates are complete, the function
signals to the scheduler that the current task instance completed its
execution so that it can schedule another task.

For instance, the function for task \lstinline!PL!  is given in
Fig. \ref{fig:code-PL}. The value returned by the external function
\lstinline!PL! is a single integer so we can directly assign its return
value to an integer variable. When the external function returns a
tuple, the output value is returned as a \lstinline!struct! pointer in
the parameters of the function. The variable \lstinline!NL_o_PL_i3!  is
the communication buffer for precedence $NL\precannot{\Fby.\Uclock
  3}PL$. It is an array of size 2 as the precedence contains a
delay. The delay is initialised at the beginning of the
function. \lstinline!PL!  alternatively reads from value 1 and from
value 0 of the array, starting with value 1
(\lstinline!NL_o_PL_i3[(instance+1)%2]!). \lstinline!PL! then copies its output value to the
communication buffer \lstinline!PL_o_order! for precedence $PL\precrel
order$. Instruction \lstinline!invoke_scheduler(0)!  signals the
completion of the task instance.

\begin{figure}[htb]
  \centering
\begin{lstlisting}[basicstyle=\scriptsize, language=C]
void *PL_fun(void * arg) {
  NL_o_PL_i3[1]=0; int instance=0;
  while (1) {
    PL_o = PL(FL_o_PL_i1,PF_o_PL_i2, NL_o_PL_i3[(instance+1)%2]);
    PL_o_order=PL_o;
    instance++;
    invoke_scheduler (0);
  }
}
\end{lstlisting}  
  \caption{Code generated for task \lstinline!PL!}
  \label{fig:code-PL}
\end{figure}

The functions for tasks \lstinline!FL! and \lstinline!NL!, which produce
data used by \lstinline!PL! are given in Fig.~\ref{fig:code-FL}. Variable
\lstinline!FL_o_PL_i1! is the communication buffer for precedence
$FL\precannot{\Dclock 4}PL$. It is updated once every 4 iterations,
(\lstinline!update_FL_o_PL_i1[instance%4]!). For precedence $NL\precannot{\Fby.\Uclock 3}PL$, \lstinline!NL!
alternatively copies its output value to the value 0 and to the value 1
of the buffer \lstinline!NL_o_PL_i3!.

\begin{figure}[htb]
  \centering
\begin{lstlisting}[basicstyle=\scriptsize, language=C]
void *FL_fun(void * arg) {
  int update_FL_o_PL_i1[4]={1,0,0,0}; int instance=0;
  while (1) {
    FL_o = FL(angle_FL_i1);
    if(update_FL_o_PL_i1[instance%4])
      FL_o_PL_i1=FL_o;
    instance++;
    invoke_scheduler (0);
  }
}

void *NL_fun(void * arg) {
  int instance=0;
  while (1) {
    NL_o = NL(NF_o_NL_i1,pos_r_NL_i2);
    NL_o_PL_i3[instance%2]=NL_o;
    instance++;
    invoke_scheduler (0);
  }
}
\end{lstlisting}  
  \caption{Code generated for tasks \lstinline!FL! and \lstinline!NL!}
  \label{fig:code-FL}
\end{figure}

The main function then creates one thread for each task, initializes the
EDF scheduler and attaches the threads to it. For instance, the thread
for task \lstinline!PL! is created by the following function call:
\lstinline!pthread_create(&tPL,&attrPL,PL_fun,!\\\lstinline!NULL)!.  \lstinline!tPL!
is the thread created for this task. \lstinline!attrPL!  contains the
real-time attributes of the task. \lstinline!PL_fun! is the function
executed by the thread. The last parameter \lstinline!NULL!  stands for
the arguments of \lstinline!PL_fun!.

\subsection{Implementing EDF with Variable Deadlines}
We choose to prototype the scheduler using \marte\ Operating System
\cite{rivas02}, which was designed specifically to ease the
implementation of application-specific schedulers while remaining close
to the POSIX model. We modify the EDF scheduler provided with the OS to
support \dwords. To summarize, the EDF scheduler is defined as a
high-priority thread created by the main function of the file. The
scheduler thread is itself scheduled by the kernel of the OS. It becomes
active only when scheduling actions must be taken, which is when the
following \emph{scheduling events} (implemented by means of signals)
occur: the current instance of a task completes its execution or a new
task instance is released. The scheduler then computes the most urgent
task among the ready tasks (tasks released and not complete yet),
resumes the execution of the corresponding thread where it stopped and
suspends the execution of the currently executing thread, if any. The
scheduler thread then becomes inactive until the next scheduling event.

The support of \dwords\ requires very few modifications. We define
\dwords\ and modify the structure describing task real-time attributes
as shown in Fig. \ref{fig:task-type}. Then, we modify the function that
programs the next instance of a task when the current instance
completes. For a task $\tau_i$, the attributes of which are described by
the value \lstinline!t_data!, the release date and the deadline of its
new instance are computed as described in Fig. \ref{fig:task-release}
($D_i[n]=R_i[n]+d_i[n]$). The function \lstinline!incr_timespec(t1,t2)!
increments \lstinline!t1! by \lstinline!t2! and the function
\lstinline!(t1,t2,t3)! sets \lstinline!t1! to \lstinline!t2+t3!.

\begin{figure}[htb]
  \centering
  \begin{lstlisting}[basicstyle=\scriptsize, language=C]
typedef struct dword { struct timespec *dds; int wsize; } dword_t;
typedef struct thread_data {
  struct timespec period;
  struct timespec initial_release;
  dword_t dword;
  struct timespec next_deadline;
  struct timespec next_release;
  int instance;
} thread_data_t;  
\end{lstlisting}
  \caption{Data type representing task real-time attributes}
  \label{fig:task-type}
\end{figure}

\begin{figure}[htb]

  \centering
\begin{lstlisting}[basicstyle=\scriptsize, language=C]
dword_t dw = t_data->dword; t_data->instance++;
incr_timespec (&t_data->next_release,&t_data->period);
add_timespec (&t_data->next_deadline,&t_data->next_release,
              &(dw.dds[t_data->instance%dw.wsize]));
\end{lstlisting}
  \caption{Releasing a new task instance}
  \label{fig:task-release}
\end{figure}

We can see that the overhead due to the support of \dwords\ remains very
reasonable. Altogether, we modified about 20 lines of code of the original
EDF scheduler, which is 300 lines of code long. We compiled and executed
the \C\ code generated for the Flight Control System and it behaved as
expected.


\section{Related Works}
\label{sec:related}
The language used in this article relies on a specific class of clocks
to handle the multi-periodic aspects of a system. Real-time periodic
clocks have also been introduced in \cite{curic05}, but they do not
include clock transformations to efficiently handle rate
transitions. Our rate-transition operators are also very similar to the
rate transition blocks of \simulink\ \cite{mathworks}. Yet, as far as we
know, for models using such blocks, the code generation tool (Real-Time
Workshop) relies on a Rate-Monotonic scheduler \cite{liu73} used with
semaphores to handle task communications, which is not an optimal
scheduling policy for this scheduling problem.

The scheduling of \textit{multi-rate} Synchronous Data Flow (SDF) is a
well studied problem (see for instance
\cite{lee87b,ziegenbein00,oh06}). In particular \cite{ziegenbein00}
studies the implementation of SDF with a dynamic scheduler using
preemption. However, though multi-rate systems are at the chore of SDF
graphs, SDF operations are \textit{not periodic}, they are not released
periodically and their relative deadline is not their period. As a
consequence, these results do not apply to our problem.

\cite{tripakis05} deals with the execution of a set of synchronous
tasks, the semantics of which is very close to our task sets, with a
dynamic scheduler. However, the authors do not detail how the
synchronous task set is obtained, for instance how it is translated from
a synchronous language, and task precedences are not specified in the
task set.

A simple solution to the problem of scheduling tasks related by extended
precedences is to unfold the extended precedence graph on the
hyperperiod of the tasks and use \cite{chetto90} to encode the simple
precedences of the unfolded graph. This solution replaces each task
$\tau_i$ by $HP/T_i$ duplicates (where $HP$ is the hyperperiod of the
tasks) in the unfolded graph. This can lead to important computation
overhead at execution as the scheduler needs to make its decisions
according to a task set that will contain many tasks. Our solution does
not duplicate any task, so the scheduler takes less time to make its
decisions.


\section{Conclusion}
We proposed a language for programming critical systems with
multiple real-time constraints, along with its compiler, which
automatically translates a program into a set of independent real-time
tasks programmed in \C\ with POSIX.13 real-time extensions. The
generated code is schedulable optimally by a slightly modified EDF
scheduler and requires no synchronization primitives. Though tasks are
scheduled concurrently and preemptions are allowed, the generated
program respects the real-time and the functional semantics of the
original program.

\bibliographystyle{eptcs}

\bibliography{fma}

\end{document}